\documentclass[12pt]{article}

\newcommand{\x}{arXiv:}
\newcommand{\m}{\mathrm}

\usepackage{graphicx}

\begin{document}
\thispagestyle{empty}
\begin{center}

\null \vskip-1truecm \vskip2truecm

{\Large{\bf \textsf{Initial Conditions For Bubble Universes}}}

{\large{\bf \textsf{}}}

{\large{\bf \textsf{}}}

\vskip1truecm

{\large \textsf{Brett McInnes}}

\vskip1truecm

\textsf{\\  National
  University of Singapore}

\textsf{email: matmcinn@nus.edu.sg}\\

\end{center}
\vskip1truecm \centerline{\textsf{ABSTRACT}} \baselineskip=15pt
\medskip
The ``bubble universes" of Coleman and De Luccia play a crucial role
in string cosmology. Since our own Universe is supposed to be of
this kind, bubble cosmology should supply definite answers to the
long-standing questions regarding cosmological initial conditions.
In particular, it must explain how an initial singularity is
avoided, and also how the initial conditions for Inflation were
established. We argue that the simplest non-anthropic approach to
these problems involves a requirement that the spatial sections
defined by distinguished bubble observers should not be allowed to
have arbitrarily small volumes. Casimir energy is a popular
candidate for a quantum effect which can ensure this, but [because
it violates energy conditions] there is a danger that it could lead
to non-perturbative instabilities in string theory. We make a simple
proposal for the initial conditions of a bubble universe, and show
that our proposal ensures that the system is non-perturbatively
stable. Thus, low-entropy conditions can be established at the
beginning of a bubble universe without violating the Second Law of
thermodynamics and without leading to instability in string theory.
These conditions are \emph{inherited} from the ambient spacetime.

\newpage

\addtocounter{section}{1}
\section* {\large{\textsf{1. Getting Inflation Started in a Bubble}}}
In string theory, the leading approach to the problem of the
cosmological constant is given by the \emph{Landscape}
\cite{kn:landscape}. String theory gives a consistent account of a
set of possible universes which are so numerous
---$\,$ $10^{500}$ is the standard estimate ---$\,$ and have values of the cosmological constant
spaced in such a way, that the value we actually observe ceases to
seem surprising. Instead we conclude that our Universe corresponds
to a point in the Landscape.

The mathematical consistency of Landscape universes does not suffice
to solve the cosmological constant problem: one needs to explain how
such a vast array of possible worlds actually comes into existence.
This is achieved by means of the nucleation of \emph{Coleman-De
Luccia bubbles} \cite{kn:deluccia}\cite{kn:frei}\cite{kn:kleban}.
These are bubbles of ``true" vacuum which spontaneously arise within
a larger spacetime containing a scalar field which is initially in a
``false" vacuum state. With a suitable potential for the scalar, and
with the usual assumptions [``potential domination"] regarding the
initial conditions for the inflaton, such bubbles can be made
compatible with the standard inflationary account of the evolution
of a universe like ours. This is the ``open Inflation" scenario
\cite{kn:lindepotential}, which works quite well in bubble universes
---$\,$ \emph{provided, of course, that Inflation can actually
begin inside a bubble:} something which is by no means obvious,
since the precise nature of inflationary initial conditions remains
to be fully understood.

Indeed, if bubble nucleation is to be taken seriously as an account
of the origin of our Universe, then it must be expected to answer
all of the long-standing questions regarding cosmological initial
conditions. In particular, it should supply answers to the following
fundamental questions:

\begin{itemize}
  \item Was the beginning singular? If not, how are the singularity theorems
  evaded?

  \item The Second Law of thermodynamics dictates that the
  Universe began in an extremely low-entropy state. How was that
  arranged? Particularly: how does one enforce the \emph{very special} conditions needed
  for Inflation to start \cite{kn:trodden}\cite{kn:albrecht}\cite{kn:couleinitial}?
\end{itemize}

The first of these questions requires no elaboration. The second
question concerns the ``specialness" of the initial conditions of
\emph{our} Universe. This specialness [or ``non-genericity"] is
still manifested, even after the passage of more than 13 billion
years, as an \emph{Arrow of time}
\cite{kn:penrose}\cite{kn:price}\cite{kn:dyson}\cite{kn:wald}. The
point is that a truly generic initial state would be dominated by
black holes\footnote{It has been argued \cite{kn:carroll} that
this is not true of a universe which begins along a non-compact
spatial hypersurface. Here we shall avoid this controversial
question by postulating that the \emph{original} universe was
created from ``nothing" \cite{kn:vilenkin}\cite{kn:ooguri} along a
\emph{compact} hypersurface. In this case, a singularity-dominated
beginning would indeed have been generic.} \cite{kn:black}. But
such an initial state would not evolve to a Universe like ours,
with its extremely strong past/future asymmetry. A crucial
instance of this is that Inflation cannot begin with such initial
conditions; in fact, as Albrecht \cite{kn:albrecht} and others
have stressed, Inflation can only begin if the inflaton is itself
initially in a very specific state, in which extremely few of the
scalar field degrees of freedom have yet been excited. If we
cannot produce a theory which \emph{necessarily entails} such
extraordinarily non-generic initial conditions for at least some
universes, then we will not be able to find a universe, even in
the Landscape, which remotely resembles our own. This was
discussed at length in \cite{kn:BBB}. [See
\cite{kn:aguirregrat2}\cite{kn:carroll}\cite{kn:arrow}\cite{kn:laura}\cite{kn:banks}\cite{kn:freese}
for various theories of the origin of the Arrow.]

We stress that settling this question is no mere technicality.
Recently it has become clear that uncertainty as to the precise
nature of the inflationary initial conditions has concrete
consequences even for the interpretation of future
\emph{observational} data. Inflation can lose its ability to predict
certain observational signatures if one weakens the usual
assumptions regarding the pre-inflationary spacetime geometry
\cite{kn:uzan}.

The bubble universe theory addresses the first of our questions in a
surprising way. As is well known, Inflation is usually not
past-eternal \cite{kn:borde}: an inflating region of spacetime must
be geodesically incomplete to the past if its average rate of
expansion is positive. A cautious interpretation of this fact
\cite{kn:guthnew} is that any inflationary history must be preceded
by some unspecified but radically different state. This is the
appropriate statement of the conclusion, because geodesic
incompleteness can have a variety of physical interpretations. The
most familiar interpretation is that incompleteness signals a
singularity, but this is not necessarily the correct interpretation
in the case of bubble universes.

Aguirre and Gratton have examined this question in the context of
their theory of the Arrow. [See \cite{kn:aguirrevaas} for a survey.]
In fact, their analysis applies quite straightforwardly also to
thin-walled Coleman-De Luccia bubbles. Their general argument
implies that these bubbles are geodesically incomplete to the past,
in the manner dictated by the Borde-Guth-Vilenkin results, simply
because the description of the bubble interior given by the
\emph{distinguished cosmological observers} inside the bubble cannot
be extended arbitrarily far into the full spacetime. In the case
where the bubble wall is infinitely thin, the explanation for this
is simple: the spacelike surfaces defined by these observers can
approach arbitrarily close to a null surface, so their volumes [or
rather the volumes of compact sets they contain] \emph{shrink
towards zero} at a finite proper time to the past of any event
inside the bubble. We shall see later that this shrinkage to zero
size actually persists even when the bubble wall is not infinitely
thin: it is a generic property of bubble interiors under certain
very mild conditions. Normally, such a situation would entail the
existence of a singularity; but it turns out that the equation of
state of the scalar field is such that zero volume \emph{does not}
imply infinite energy density. Thus, in this case, the
incompleteness signals that the bubble spacetime is extensible,
\emph{not} singular; it can be extended, via the bubble wall, into
the ambient spacetime. This is how the bubble universe proposal
deals with the singularity problem.

This solution of the singularity problem can be questioned: clearly
it depends on very strong assumptions about the exact matter content
of the pre-nucleation spacetime. If, for example, the ambient
spacetime contains other fields or objects, one will need to
investigate their effects if they are absorbed by the bubble and
encounter the zero-volume spatial slice; also, quantum effects [such
as the Casimir effect] may alter the classical picture of the matter
content in a decisive way. In the case of an infinitely thin bubble
wall, incursions by external objects could be disastrous. The most
dramatic example of such incursions involves the collision of a
bubble with \emph{another} bubble. The problem of understanding
these questions in that context is currently the subject of
intensive investigation; see the detailed discussions given in
\cite{kn:aguirregrat}\cite{kn:aguirrejoh}.

Leaving these complications aside for the moment, we can summarize
by saying that the simplest versions of bubble universes offer an
approach to the singularity problem simply by arguing that the
earliest form of matter [necessarily] had an unusual equation of
state, such that its energy density was not related to spatial
volume in the familiar way. This permits an interpretation of the
Borde-Guth-Vilenkin results in a way that does \emph{not} involve
singularities.

An answer to the first of our questions should set the scene for an
answer to the second: since the zero-volume state\footnote{Strictly,
we should say that the spatial sections have volumes which can be
made \emph{arbitrarily small}, not exactly zero; the distinction is
however not important here.} at the beginning of the bubble universe
is \emph{not} singular, \emph{there is no obstruction to relating
the thermodynamic conditions in the early bubble universe to
conditions in the ambient spacetime}.

Thus, the problem of cosmological initial conditions can only be
addressed, in the bubble universe context, by applying the Second
Law of thermodynamics to the bubble nucleation process. In this work
we argue that this suggests a small but significant modification of
the usual approach to bubble nucleation theory. The idea that even
exponentially suppressed corrections to the Coleman-De Luccia
instanton can be important has been advocated by Buniy, Hsu, and Zee
\cite{kn:hsu}; here we consider much less drastic modifications,
which alter the bubble geometry only at the very earliest [bubble]
times.

General aspects of applying the Second Law to bubble nucleation are
explained in Section 2. The key point here is that, for a bubble
universe to resemble our own, its initial total entropy must be low
as seen by the \emph{distinguished observers} inside the bubble
---$\,$ the observers to whom the spatial geometry appears to be
isotropic. But this is very difficult to arrange, because these same
observers are the ones whose spatial sections shrink to zero volume
as they probe backwards in time, and small spatial sections have a
\emph{very strong tendency to be anisotropic}. This key point will
be reviewed in some detail. We argue that the most natural ---
though perhaps not the only ---$\;$ way of avoiding this problem is
to find some means of avoiding a zero-volume ``initial" state for a
bubble universe.

In Section 3, we discuss in detail the way in which a standard
Coleman-De Luccia bubble universe avoids being [initially] singular
and develops an Arrow of time. We focus first on the case of
negative vacuum energy inside the bubble, since the points we are
making can be seen most clearly in that case [which does occur in
the Landscape, in the form of ``terminal vacua"]. Using singularity
theory, we show that the zero-volume spatial section will also be
present in the case of positive vacuum energy. It can only be
avoided by modifying the bubble universe in a way that violates the
\emph{Null Ricci Condition} or NRC. This is the statement that the
Ricci tensor satisfies
\begin{equation}\label{NEC}
\m{R_{\mu\nu}\,n^{\mu}\,n^{\nu}\;\geq\;0}
\end{equation}
at all points in spacetime and for all null vectors $\m{n^{\mu}}$;
it is equivalent to the \emph{Null Energy Condition} or NEC in cases
where corrections to the Einstein equations can be neglected. If
such corrections are important then the NRC can be violated even if
the NEC is satisfied; in such a case we may speak of
\emph{effective} violations of the NEC\footnote{That is, violation
of the NRC amounts to violating the NEC for the effective
stress-energy-momentum tensor obtained by absorbing the corrections
into the physical stress-energy-momentum tensor. Our main example
[the Casimir effect] involves ``true" NEC violation, but the
distinction being made here is important, and should be borne in
mind; see for example \cite{kn:bojowald}.}. Our conclusion is that
the required modification violates the NEC, though perhaps only
effectively.

Real and effective NEC violations in string cosmology have been
discussed in
\cite{kn:unstable}\cite{kn:tallandthin}\cite{kn:singularstable}, and
have recently attracted much more interest
\cite{kn:ovrut}\cite{kn:cremi}. The perennial concern with regard to
NEC violation is the possibility that it might lead to some kind of
fatal instability \cite{kn:carrtrod}. Arkani-Hamed et al.
\cite{kn:nimah1}\cite{kn:nimah2} argue that NEC violation is not
acceptable in string theory \emph{except} when it is global and
quantum-mechanical, as in the case of the \emph{Casimir effect}
\cite{kn:coule}, or in other very special conditions [such as those
associated with orbifold planes]. However, even in these cases one
must also take into account certain \emph{non-perturbative} string
effects, because it has been shown that these frequently do lead to
problems when NEC violation occurs. In particular, we have to take
into account the brane-antibrane pair-production instability
analysed by Seiberg and Witten \cite{kn:seiberg} and subsequently by
Maldacena and Maoz \cite{kn:maoz} and by Kleban et al.
\cite{kn:porrati}. This instability means that NEC violation ---$\,$
even if it is only \emph{effective} ---$\,$ is \emph{not} always
physically acceptable even in the cases where it does not lead to
problems at the perturbative level.

In Section 4, we examine a particular model in which the interior of
a bubble universe begins, with the aid of the Casimir effect, along
a surface of non-zero minimal volume. We are able to show that,
despite the violation of the NEC entailed by Casimir energies, the
spacetime narrowly avoids becoming unstable in the Seiberg-Witten
sense. Thus we have a toy model of a bubble universe which has
satisfactory initial conditions for Inflation; it is able to inherit
an Arrow of time.

We stress that the metric we find is asymptotic to one of the
metrics normally used to describe bubble interiors, and differs
substantially from such a metric only for an extremely short time.
Thus our conclusions do not invalidate the large recent literature
on eternal Inflation in any way; nor, of course, are we suggesting
that there is anything erroneous in the original Coleman-De Luccia
analysis. The objective is simply to show that the bubble universes
that populate the Landscape can in fact have initial conditions
similar to those of our own Universe.

\addtocounter{section}{1}
\section* {\large{\textsf{2. Bubble Nucleation Respects The Second Law}}}

In this section we construct a very general argument to the effect
that the Second Law of thermodynamics has specific consequences for
the spatial geometry of the very earliest phase of a bubble
universe.

First, note that, unlike the baby universes considered by Farhi and
Guth \cite{kn:fargu} [see also \cite{kn:aguirrejohn}], a Coleman-De
Luccia bubble is \emph{not} isolated from the original spacetime: on
the contrary, the bubble expands into the ambient universe and is
permanently exposed to signals from it. Indeed, to a family of
observers inside a bubble which nucleates in an approximately
Minkowski spacetime, \emph{the entire exterior spacetime lies to the
past}. The Second Law of thermodynamics now has the following major
consequence: \emph{we cannot simply ``re-set" the initial conditions
inside the bubble to suit ourselves.} The initial thermodynamic
state of a bubble is set by the outside conditions and by what
happens as one moves through the wall. In this connection, one
should not expect the bubble wall to preserve all highly-ordered
structures it encounters
---$\,$ let alone generate them. That is, passage through the wall could
lead to a dramatic increase of certain kinds of entropy. This is
consistent with Coleman and De Luccia's description of passing into
such a bubble as ``the ultimate ecological catastrophe".

It follows from these simple observations that, if the bubble
interior has extremely low initial entropy, this can only be a
result of \emph{inheriting} that condition from the ambient
spacetime. Answering our question then amounts to establishing the
following two statements.

\begin{itemize}
  \item The ambient spacetime had extremely low
entropy.

  \item The inevitable increase in the entropy caused by bubble nucleation does not
  appear to be large as seen by an internal observer: low entropy is heritable.

\end{itemize}

One way of approaching the first point was proposed in
\cite{kn:aguirregrat2}; in \cite{kn:arrow} we addressed it in a
different way, by arguing that bubble universes nucleate in a
``mother universe" which itself is the result of ``creation from
nothing", after the manner of Vilenkin \cite{kn:vilenkin} and Ooguri
et al. \cite{kn:ooguri}. With a suitable spatial topology, one can
use deep theorems from global differential geometry to argue that
the original universe necessarily had a perfectly [locally]
isotropic initial spatial section. This means that the initial
\emph{gravitational} entropy\footnote{The concept of gravitational
entropy has not yet been made entirely precise: see for example
\cite{kn:tod}\cite{kn:ram}. That gravitational systems behave
consistently with the Second Law is nevertheless not in doubt, and
this is all we need here.} was [necessarily] as low as possible, and
indeed this is precisely why the \emph{total} entropy of this
initial universe was low \cite{kn:penrose}\cite{kn:price}: extreme
isotropy rules out black holes, which would otherwise strongly
dominate the entropy accounting in a spatially compact universe. The
gravitational entropy then increases due to the usual inflationary
fluctuations, which mar the perfect geometric regularity of the very
earliest spatial sections
---$\,$ if only to a microscopic degree.

This brings us to the second point. The Second Law dictates that the
gravitational entropy cannot decrease during the bubble nucleation
process. The question now is: what form will the increase take,
\emph{as seen by interior observers}? It is important to understand
here that the bubble interior differs, in one crucial particular,
from Minkowski or [anti] de Sitter spacetime. These latter
spacetimes have very large [in fact, maximal] isometry groups,
corresponding to their extremely simple matter contents. It follows
that they do \emph{not} have distinguished families of observers, as
a generic FRW cosmological model does. Thus, for example, [regions
of] de Sitter spacetime can be foliated in many different ways by
spacelike surfaces having a variety of intrinsic geometries, and
none of these foliations has a preferred status; for all of them
correspond to observers who \emph{see the same thing}, namely
isotropic dark energy with a particular invariant energy density. By
sharp contrast, the surfaces of approximately constant scalar energy
density inside the bubble \emph{do} distinguish a special class of
observers. \emph{These} observers are the ones who, using whatever
coarse-graining they find appropriate, must deduce very low-entropy
conditions in \emph{their} earliest history, if the bubble universe
is to have an Arrow of time. This is another sense
---$\,$ apart from the ``ultimate ecological catastrophe" aspect
---$\,$ in which the interior of a bubble universe is \emph{not}
analogous to [say] the interior of a forward light-cone in de Sitter
spacetime. The bubble universe contains distinguished observers
whose [coarse-grained] observations are what we have to explain.

We begin our investigation of this question by noting that, for
Inflation to start, what is really needed is low
\emph{gravitational} entropy: if the spatial sections are too
irregular, this will not be consistent with the required initial
conditions for the inflaton. \emph{Other} forms of entropy, such as
the entropy of the Gibbons-Hawking radiation \cite{kn:gibhawk}
associated with a cosmological horizon, will actually increase
substantially during bubble nucleation, but this will not interfere
with the inflaton initial conditions. [Nor, however, will it help to
establish the particular form we need for these conditions.] This
observation refines our question considerably.

Now as we have seen, the characteristic property of the earliest
spatial sections inside the bubble is that their volume scales are
arbitrarily small; this is the proper interpretation of the
Borde-Guth-Vilenkin results \cite{kn:aguirrevaas}. \emph{But one
does not expect ``small" spatial sections to correspond to low
gravitational entropy.} This can be explained as follows. Just as
Inflation leads an observer to think that his spatial sections have
become smoother\footnote{See \cite{kn:gibsol} for the precise
statement of ``cosmic baldness", and \cite{kn:BBB} for a
discussion.}, so also an unlimited \emph{contraction} of a spatial
section will make any irregularities more and more apparent. To put
this another way, suppose that we consider the history of a small
spatial patch in the present Universe. As we trace it back in time,
we will see it becoming less and less isotropic around a generic
point.

To see this, one needs to study the effect of including anisotropy
in the spacetime dynamics. The anisotropy contributes a term of the
form C/$a(t)^6$ to the field equations, where $a(t)$ is the scale
factor and C is a constant. [This is explained extremely clearly in
\cite{kn:turok}, which should be consulted for the
details\footnote{Generically these irregularities are of the kind
originally discussed by Belinsky, Khalatnikov, and Lifschitz
---$\,$ see \cite{kn:uggla}\cite{kn:imponente} for recent detailed
discussions.}.] This means that, as we go back in time, the
anisotropy grows much more rapidly than the energy density of
ordinary matter and radiation [or of any kind of dark energy], so it
dominates the dynamics if the sections become sufficiently small, at
least in the absence of very exotic forms of matter.

This discussion explains why one does not expect a \emph{realistic}
zero-volume state, such as that of a Big Crunch or a black hole, to
be geometrically regular, though of course the intrinsic geometry of
the late spatial sections is very regular in \emph{idealized} FRW or
Schwarzschild spacetimes. \emph{Generically, spatial sections with
very small volume scales correspond to high gravitational entropy}.
It follows that observers inside a bubble universe, having deduced
from observations that Inflation took place, will have to conclude
that the initial conditions for Inflation were made possible by an
\emph{infinite} fine-tuning at the zero-volume state --- infinite in
the sense that the C/$a(t)^6$ term can only be ignored, if $a(t)$
really vanishes, if C is set \emph{exactly} equal to zero. If they
are aware that they live in a bubble, they will be forced to
conclude that they owe their existence to a massive
violation\footnote{The Second Law, being statistical, can of course
be ``violated"; but the dire consequences of assuming that the
current status of our Universe can be explained in that way are well
known: see \cite{kn:linde}\cite{kn:page}\cite{kn:carlip}, and
\cite{kn:gott} for a survey.} of the Second Law. [They will realise
that the Gibbons-Hawking entropy has increased, but since anisotropy
generically completely dominates the dark energy density, they will
not be able to explain the situation by using this fact.]

There are basically four ways to deal with this problem. The first
is to postulate the presence of some kind of matter with an energy
density that grows even more rapidly, as volumes shrink towards
zero, than the anisotropy; for example, a scalar field with a very
negative potential \cite{kn:turok}\cite{kn:podolsky2}. One might
then try to arrange for the growth of the entropy to be diverted
away from the spatial geometry and into the scalar field. While this
idea deserves [and requires] further development, it does not appear
to be compatible with the Landscape picture and we shall not
consider it further.

The second approach is the usual one: we simply ignore the effects
of its environment on the bubble, and use idealized models of the
geometry. Recently, however, it has been recognised
\cite{kn:gagv}\cite{kn:freihorshe}\cite{kn:aguijohsho}\cite{kn:fish}\cite{kn:worldscollide}\cite{kn:aguirrejoh}
that \emph{collisions} of bubbles are of the utmost importance,
since, even if Inflation is able to start in the aftermath, a
permanent ``memory" of the collision may be retained by the bubbles.
Furthermore, the collisions release radiation into the ambient
spacetime \cite{kn:worldscollide}. However, bubble collisions are
just the most dramatic way in which the ambient spacetime can affect
the initial conditions of a bubble universe. On a vastly smaller
scale, the inflaton field in the ambient spacetime will suffer
scalar and tensor perturbations; these may be tiny, but they must
have \emph{some} effect on the geometry of the bubble interior if
they strike or are absorbed by the bubble. To suppose otherwise
would, once again, amount to a violation of the Second Law of
thermodynamics. These developments render obsolete the picture of a
bubble existing in splendid isolation; we now have to think in terms
of a bubble expanding in an environment where \emph{it is constantly
subjected to a bombardment of external signals of greater or lesser
degrees of intensity.} Since the bubble initial conditions are
``fine-tuned" [in the sense discussed earlier], it is hard to see
how to justify ignoring these signals.

In a third approach, one might accept that external signals have
these effects at a \emph{generic} point on the earliest spatial
slices of the bubble universe, but try to argue that there will
always be \emph{some} extremely atypical regions which remain
undisturbed\footnote{An interesting variant of this argument, to the
effect that the region near to the centre of the bubble is
particularly favoured, will be mentioned in the next section.}. The
observed Universe might have evolved from a tiny patch of this sort.
This amounts to an invocation of the anthropic principle. Rather
than become involved in the anthropic debate, we note instead that
it is generally accepted that alternatives to that approach should
always be fully investigated.

The fourth approach is the one to be explored here: we can try to
\emph{prevent} the spatial sections inside the bubble from ever
being too small. This has to be done by considering small
modifications of the Coleman-De Luccia analysis, taking into account
effects previously neglected. By considering sub-leading corrections
to the Coleman-De Luccia instanton, as advocated by Buniy et al
\cite{kn:hsu}, one can study tunnelling processes not described
purely by an analytic continuation of a single Euclidean solution.
This opens the way to including global quantum effects, similar to
the Casimir effect \cite{kn:coule}. These will \emph{not} change the
standard large-scale picture of the bubble spacetime; they are only
important when the bubble interior is very small. One can think of
this procedure in terms of allowing other forms of energy, in
addition to that of the scalar field, to act on the spacetime
geometry.
\begin{figure}[!h]
\centering
\includegraphics[width=0.8\textwidth]{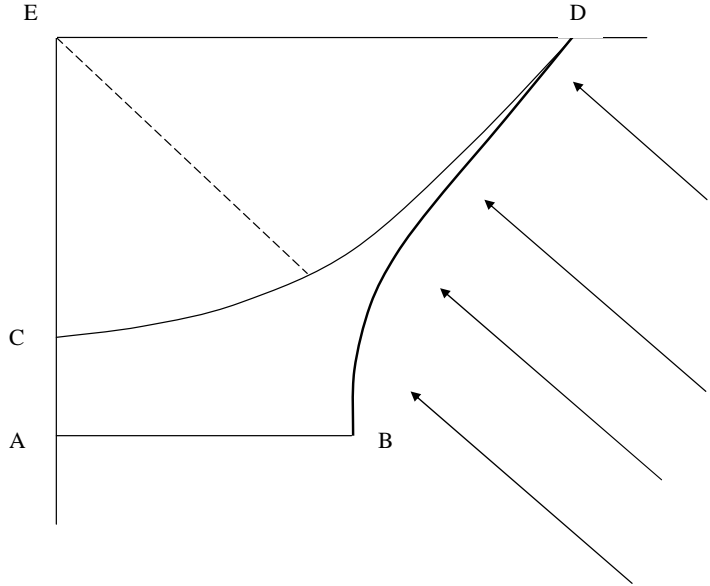}
\caption{Positive Vacuum Energy Bubble in Positive Vacuum Energy
Background.}
\end{figure}

A schematic Penrose diagram of the nucleation of a
positive-vacuum-energy bubble residing in a larger spacetime [which
itself has positive vacuum energy] is given in Figure 1. The bubble
nucleates along AB and the outer surface of its wall is represented
by BD. Notice that the entire initial spatial section inside the
bubble is exposed to outside influences. A gravitational wave [say]
in the outside world can reach \emph{any} point in the initial
spatial section of the bubble even if it originates from a point
deep inside the de Sitter ``bulk"; this is symbolized by the arrows
in the diagram.

Following Aguirre and Gratton \cite{kn:aguirregrat2}, we think of
semi-classical bubble nucleation as a three-stage process: the
ambient spacetime and the bubble interior [ECD in the diagram] can
be described more or less accurately by classical geometry, but the
transition region [ACDB in the diagram] is a predominantly quantum
domain. The idea is that, in that domain, quantum effects prevent
the characteristic geodesic focussing associated with classical
gravity, thus ruling out anything analogous to a shrinking of
spatial sections to zero size. This is a reasonable expectation,
because it is known that this ``quantum de-focussing" is precisely
what happens in the course of the Hawking evaporation of a black
hole --- see \cite{kn:strominger} for a particularly clear
discussion of this.

While this fourth proposal seems to be the simplest way of
explaining how a bubble universe can inherit the geometric
regularity of the ambient spacetime, it forces us to confront a
basic issue: what, exactly, \emph{are} the initial conditions for
the bubble universe? That is, if we consider the spatial section
along which a semi-classical description first becomes appropriate,
the surface CD in Figure 1, we need to know something about the
conditions imposed on this section by the quantum domain. Without
this information, we will of course be unable to predict the
subsequent evolution of the bubble interior.

The appropriate initial [and boundary] conditions for matter
fields will be discussed in Section 4; for the moment, we shall
focus on the initial conditions for the spacetime geometry of the
bubble. We propose that the correct initial condition for the
interior semi-classical spacetime is that the ``initial" spatial
section CD is of \emph{minimal} but non-zero volume [specifically,
that it is a spacelike surface of [approximately] \emph{vanishing
extrinsic curvature}].

There are four reasons for thinking that this is the right
procedure. First, one can argue that, in string theory, it is not
reasonable for \emph{any} cosmological model with compact spatial
sections to have spatial volumes much below the cube of the string
length scale; so there should be a spatial section of minimal volume
or zero extrinsic curvature, and that spatial section is the natural
locus for a semi-classical description to be appropriate
\cite{kn:BBB}. [Strictly speaking, the spatial sections of a bubble
universe are infinite in extent, but in the ``holographic"
interpretation we adopt here \cite{kn:bouff} they are
\emph{effectively} finite. This will be discussed later.]

Second, a connection between zero extrinsic curvature and low
entropy is suggested by Verlinde's \cite{kn:verlinde} observation
that Cardy's formula for the entropy of a conformal field theory can
reproduce the Friedmann equation. Minimal volume is then naturally
associated with low ``holographic entropy", because the latter is
related \cite{kn:huang1}\cite{kn:huang2} to the extrinsic curvature
of spatial sections.

Third, the Borde-Guth-Vilenkin theorem implies that the only way
an inflating spacetime can avoid having zero-volume spatial
sections is to have a longer history of contraction than of
expansion. We therefore need to use \emph{part of} a ``bouncing"
cosmological model \cite{kn:novello}, that is, part of a spacetime
which does have a spacelike surface of zero extrinsic curvature.
This surface is the only distinguished one in the spacetime, and
so it is natural to use the part which begins along this surface.
[True ``bounce" cosmologies, including the contracting part, are
interesting \cite{kn:piao}\cite{kn:peter}, but they encounter
notorious entropic difficulties of precisely the kind we hope to
resolve here, and the most recent work \cite{kn:couleholo} only
serves to reinforce these doubts regarding the thermodynamics of
``bounces".] To put it another way: if the surface CD in Figure 1
does \emph{not} have vanishing extrinsic curvature, then this
non-zero object would define a new fundamental time scale. [In the
special case of FRW cosmology, the extrinsic curvature is given by
$-\,\dot{a}(t)/a(t)$, where $a(t)$ is the scale factor and the dot
denotes a \emph{time} derivative.] It is hard to see how such a
scale could arise in string theory.

Finally, our picture of the origin of the Arrow of time in the
ambient spacetime \cite{kn:arrow} supposes that the latter emerges
from a state with no classical description ---$\,$ that is, from
``nothing" \cite{kn:vilenkin}\cite{kn:ooguri} ---$\,$ along a
surface of zero extrinsic curvature; so, to be consistent, we should
assume that a similar principle applies when classicality emerges
inside the bubble.

With a concrete proposal for the initial conditions of a bubble
universe, we can explore the structure of the spacetime in the early
history of such universes. Before doing so, however, let us see more
concretely how all of these observations apply to the usual
description of bubble universes.

\addtocounter{section}{1}
\section* {\large{\textsf{3. The Arrow of Time in Standard Bubble Universes}}}
The original examples of bubble universes were those studied by
Coleman and De Luccia \cite{kn:deluccia}, who showed that they arise
in the interior of bubbles of true vacuum nucleating in a false
vacuum defined by a local minimum of a scalar field potential
V($\varphi$). The scalar is assumed to be the \emph{only} form of
matter present in the spacetime. Let us consider in detail how an
Arrow can arise, under this assumption.
\begin{figure}[!h] \centering
\includegraphics[width=0.8\textwidth]{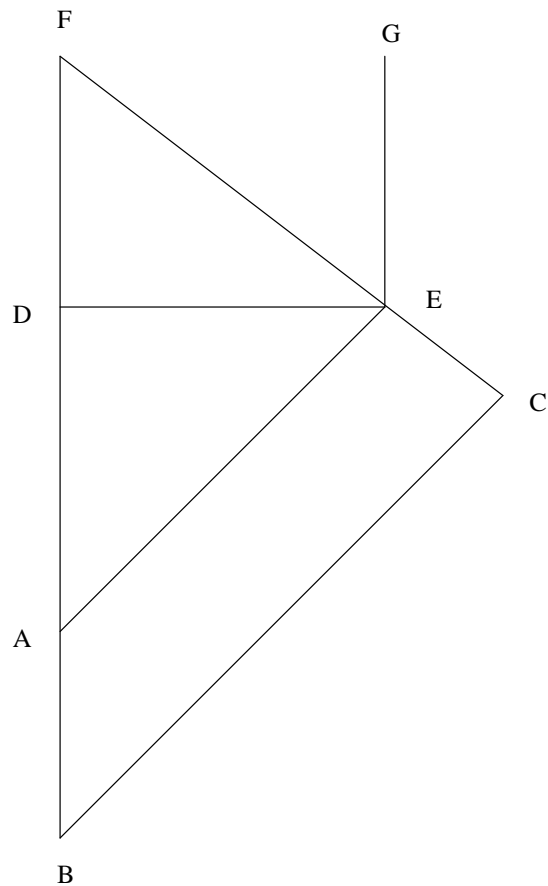}
\caption{Negative Vacuum Energy Bubble in Minkowski Spacetime.}
\end{figure}
Let us begin with a bubble of \emph{negative} vacuum energy
nucleating in a Minkowskian background; that is, following Coleman
and De Luccia, we represent the scalar field inside the bubble by a
negative cosmological constant, and we treat the wall as being
infinitely thin. The conformal geometry of the bubble and its
environment is depicted in Figure 2.

The original Minkowski space is represented by the triangle BCF, and
the bubble wall is the null surface AE. In a more realistic version,
the bubble wall would be timelike; but then the wall would
accelerate, so the curve representing it would not be geodesic and
so it is still able to terminate on future null infinity. This is
important, because it means that the bubble is exposed to signals
from the entire exterior spacetime, whether the wall is thin or not.
The region FAEG represents the interior of the bubble; it is a part
of the maximally symmetric simply connected four-dimensional
spacetime of negative vacuum energy density $-\,1/8\pi$L$^2$, the
anti-de Sitter spacetime AdS$_4$.

The timelike conformal boundary of this part of AdS$_4$ is
represented by EG; as usual this causes the surface DE to have a
future \emph{Cauchy horizon}, EF, and also a past ``Cauchy horizon",
AE. [That is, AE is a Cauchy horizon in the original AdS$_4$.] The
region FAE can be covered by coordinates such that DE is t = 0; in
these coordinates the metric in FAE takes the form
\begin{eqnarray}\label{eq:A}
g(\mathrm{AdS_4}) \; =\;
\m{-\,dt^2}\;+\;\m{cos^2(t/L)}\,[\mathrm{dr^2\; +\;\m{L^2}\,
sinh^2(r/L)}\{\mathrm{d}\theta^2 \;+\;
\mathrm{sin}^2(\theta)\,\mathrm{d}\phi^2\}].
\end{eqnarray}
Notice that the spatial sections are, at least locally\footnote{That
is, ignoring topological questions.}, copies of the hyperbolic
space, H$^3$. Notice too that the spacelike hypersurfaces near to t
= $\mp\,\pi$L/2 have volumes which are tending to \emph{zero}. These
regions are, respectively, the past and future Cauchy horizons of
DE, that is, they correspond to AE and EF in the diagram. They are
\emph{not} singularities in the sense of having divergent curvature,
despite the fact that their volume vanishes, because the equation of
state of vacuum energy
---$\,$ the only form of energy in pure AdS$_4$
---$\,$ has a particular form: the energy density is constant, and
cannot diverge under any circumstances. Notice finally that the
spacetime is apparently time-dependent; this is due to the fact that
the corresponding inertial observers are not the Killing observers:
but there is in fact a timelike Killing vector in this spacetime.
Time \emph{does not pass} in this bubble; vacuum energy cannot
``age", and the spacetime itself is static.

Of course, this model is unrealistic in several ways: the bubble has
a thin wall, the scalar field is treated as if it were exactly
equivalent to vacuum energy, and the computation assumes strict
semi-classical dominance of tunnelling amplitudes \cite{kn:hsu}; in
particular, no allowance has been made for any kind of perturbation
impinging on the bubble wall from outside.

In fact, a first step towards greater realism was taken by Coleman
and De Luccia themselves, who gave a beautifully simple discussion
of the consequences taking into account the first-order corrections
to the thin-wall approximation. The essential point is that while
the rotational symmetry of their instanton continues to enforce the
initial vanishing of the time derivative of $\varphi$, it cannot
force $\varphi$ itself to vanish exactly. The results can best be
pictured in the following way. We remarked above that the metric in
equation (\ref{eq:A}) \emph{appears} to represent a spacetime, with
H$^3$ spatial sections, which is dynamic. This is not in fact the
case. Coleman and De Luccia find, however, that the slightest
perturbation away from the exact thin-wall conditions produces a
spacetime which \emph{really is} dynamic: the Killing vector is
lost. This paves the way for an Arrow to be established.

There is a crucial point here, however: making the spacetime more
realistic can be expected to do away with the Cauchy horizons of the
exact AdS$_4$ spacetime, since these horizons are typically
unstable. But the Cauchy horizons do not simply disappear:
generically, \emph{they turn into spacelike surfaces of zero
volume.} This aspect of AdS$_4$ was understood long ago: see for
example the discussion on page 172 of the review article of Tipler,
Clarke, and Ellis \cite{kn:tiplerreview}. One says that the Cauchy
horizons are replaced by \emph{crushing singularities}, though these
need not be ``singular" in the sense we use here; see
\cite{kn:tiplerreview}, page 166, for a definition.

With all this in mind, we proceed to the usual description of
\emph{realistic} versions of bubble universes with negative vacuum
energy; it runs as follows. Once we recognise that the spacetime is
dynamic, we must expect the scalar field to fluctuate and to
transfer energy to any other field to which it may be coupled, as
happens in inflationary re-heating. By the time the bubble interior
nears the zero-volume spatial section which replaces the Cauchy
horizon EF in Figure 2, then, it will contain forms of matter with
\emph{conventional} equations of state, such that the energy density
\emph{does} diverge at late times. In short, a realistic version of
a negative-energy bubble terminates in a true Crunch. These are the
``terminal vacua" in the Landscape.

But if we grant that one Cauchy horizon becomes singular, why does
that not happen along the other Cauchy horizon, AE in Figure 2? To
see how this works, we consider the situation described in
\cite{kn:frei}, where the scalar field inside the bubble is no
longer represented by a simple vacuum energy.

The metric is a FRW metric with spatial sections of constant
negative curvature; the O(4) symmetry group of the Coleman-De Luccia
instanton becomes the O(1,3) group of [\emph{local}] symmetries of
three-dimensional hyperbolic space. Regularity of the Euclidean
instanton, which has a characteristic length scale L, guarantees
that the Lorentzian metric has the following general form:
\begin{eqnarray}\label{eq:EE}
g(\mathrm{Bubble }) \; =\;
\m{-\,dt^2}\;+\;\m{a(t/L)^2}\,[\mathrm{dr^2\; +\;\m{L^2}\,
sinh^2(r/L)}\{\mathrm{d}\theta^2 \;+\;
\mathrm{sin}^2(\theta)\,\mathrm{d}\phi^2\}],
\end{eqnarray}
where, if we choose zero as the origin of time,
\begin{equation}\label{eq:SMALLA}
\m{a(t/L) \;=\; t/L\;+\;\textbf{O}(t^3/L^3).}
\end{equation}
In a FRW cosmology with negatively curved spatial sections [with
curvature proportional to $-$1/L$^2$], one can show
straightforwardly that the pressure is given by
\begin{equation}\label{eq:PRESSURE}
\m{8\pi p \;=\;{{1/L^2\;-\;2a\ddot{a}\;-\;\dot{a}^2}\over{a^2}},}
\end{equation}
where the dot denotes a proper time derivative, while the energy
density is
\begin{equation}\label{eq:DENSITY}
\m{8\pi \rho \;=\; 3{{\dot{a}^2\;-\;1/L^2}\over{a^2}}.}
\end{equation}
Applying this to the case at hand, we find, if we set
$\m{a(t/L)\,\approx \, t/L \,+\,\alpha t^3/L^3}$, where $\alpha$ is
a constant [which is negative if the energy density is negative at
small t], that
\begin{equation}\label{eq:PRESSUREAPPROX}
\m{8\pi p \;\approx\; {{-\;18\alpha\;-\;21 \alpha^2
t^2/L^2}\over{L^2(1\;+\;\alpha t^2/L^2)^2}}}
\end{equation}
and
\begin{equation}\label{eq:DENSITYAPPROX}
\m{8\pi \rho \;\approx\; {{18\alpha\;+\;27 \alpha^2
t^2/L^2}\over{L^2(1\;+\;\alpha t^2/L^2)^2}}.}
\end{equation}
Thus we see that, even though the spatial sections shrink to zero
size as t approaches zero, neither the pressure nor the density
diverges in this limit, as would be the case for any normal form of
matter or radiation. The scalar field pressure and density are
related by a somewhat bizarre equation of state [obtained by
eliminating t in the above relations for p and $\rho$], and this is
what allows this field to avoid causing a singularity at t = 0.

Obviously this bubble universe has a very definite Arrow of Time: it
begins in a perfectly smooth non-singular state and ends in a [no
doubt highly irregular] Crunch singularity. In particular, it is
clear that the gravitational entropy is initially low and finally
very large. But the origin of this Arrow is all too clear: \emph{we
built it in}, by \emph{assuming} that the tunnelling originated in
perfectly smooth Minkowski space, which justifies the description of
the tunnelling by a perfectly smooth, exactly O(4)-invariant
Euclidean instanton. We have in fact been guilty of practising
Price's \cite{kn:price} ``\emph{double standard}": we made
assumptions about the beginning of the bubble universe that we would
never apply to its ``generically" singular end. If we had allowed
for perturbations in the ambient spacetime propagating into the
bubble and ---$\,$ in accord with the Second Law ---$\,$ disturbing
the geometry there, then, as we discussed in Section 2, the
\emph{arbitrarily} small spatial slices near to t = 0 would not be
perfectly smooth, and the scalar field might not be in a
sufficiently low-entropy initial state.

It is true that, even in this case, the entropy of the initial state
could still be \emph{somewhat} lower than that of the final state,
so the bubble might have an arrow of a sort. This argument has
particular force when we consider the region of spacetime near to
the initial nucleation event. Recall that Aguirre and Gratton
\cite{kn:aguirrevaas} argued that the geodesic incompleteness of a
[thin-walled] bubble universe is due to the fact that the spatial
sections defined by distinguished observers have a tendency to
become null. This tendency is less marked near the centre of the
bubble, and so one might hope that the growth of anisotropies as one
moves back in time could be controlled in that region. If this is
true, then it might pave the way towards dealing with the notorious
problems associated with the infinite extent of the bubble spatial
sections, since only a relatively small region of the bubble could
come to resemble our Universe.

Against this, however, one has to bear in mind that the only
Universe we have observed does not just have ``low" initial entropy:
its initial entropy is \emph{fantastically} lower than it might have
been, as Penrose \cite{kn:penrose} has shown by a well-known
calculation. Thus our task is not just to show how bubble universes
can have \emph{some} kind of Arrow --- rather, we have to show how
they can have an Arrow \emph{of the kind we observe}. Again, it is
hard to see how such delicate initial conditions can be maintained
in the face of anisotropies which, as we have discussed, grow
extremely rapidly if spatial sections are allowed to become
arbitrarily small. This is a matter which can be settled only by
means of a detailed calculation, which we shall not attempt here.

Because it has both a beginning and an end, the negative-energy
bubble is particularly suited to a discussion of the Arrow, but the
problem persists even in the more directly interesting case of a
bubble with positive vacuum energy. An Arrow of time will emerge in
this case too, provided that the scalar field is in a sufficiently
low-entropy state initially. But in this case too we will find a
geometry like the one given in equations (\ref{eq:EE}) and
(\ref{eq:SMALLA}), with a zero-volume initial state. Again, the
pressure and density [given by (\ref{eq:PRESSUREAPPROX}) and
(\ref{eq:DENSITYAPPROX}), but with positive $\alpha$] do not diverge
even at zero volume, but the problem of large initial gravitational
entropy persists. For that problem is associated with zero volume,
not with the question as to whether a singularity is present.

As mentioned above, one way to deal with this problem would be to
investigate, in detail, whether the growth of anisotropies can be
controlled in the favourable region near the centre of the bubble.
Here, instead, we shall postulate that the scalar field is
\emph{not} the only important contributor to the total energy
density inside the bubble: there must be another contribution due to
quantum effects which ``de-focus" geodesics, so that there is never
any surface of zero volume. The question now is: what is the
mathematical description of this quantum contribution to the energy
density?

Since the geometry of the earliest spatial sections inside a bubble
universe is precisely the issue here, we must not base our arguments
on FRW spacetime geometries. It will be useful, however, to begin by
reminding ourselves of the reasons for the fact that FRW models with
negatively curved spatial sections tend to be geodesically
incomplete. The relevant singularity theorem is the one due to
Penrose; it may be stated as follows. [See \cite{kn:waldbook}, page
239 for the theorem and the relevant concepts].

\bigskip
\noindent \textsf{THEOREM [Penrose]: Let M$_4$ be a spacetime
satisfying the Einstein equations and the following conditions:}

\noindent \textsf{[a] The Null Ricci Condition [NRC] holds.}

\noindent \textsf{[b] M$_4$ is globally hyperbolic and contains a
non-compact Cauchy surface.}

\noindent \textsf{[c] M$_4$ contains a trapped surface.}

\noindent \textsf{Then there is at least one incomplete
future-directed null geodesic orthogonal to the trapped surface.}

\bigskip

With regard to condition [a], recall the discussion of the NRC in
Section 1; with regard to [b], note that ``non-compact Cauchy
surface" can be weakened to ``Cauchy surface with a non-compact
universal cover". Thus, compactifying the hyperbolic spatial
sections of a bubble\footnote{That is, projecting to a compact
quotient of hyperbolic space by a discrete freely acting group of
isometries.} does not \emph{in itself} allow us to avoid geodesic
incompleteness here ---$\,$ but see Section 4, below.

Assuming that the NRC is not violated, the only condition of this
theorem which needs to be verified in the case of FRW cosmologies
with negatively curved spatial sections [whether compactified or
not] is the last. Take the metric given in equation (\ref{eq:EE})
and consider a 2-sphere with radial coordinate r at time t; its area
is 4$\pi$L$^2$a(t)$^2$sinh$^2$(r/L). The orthogonal
\emph{outward}-directed set of past-pointing null geodesics
intersect the surface t = t + dt [with negative dt] at radial
coordinate r $-$ dt/a(t), and so the change in the area of the
sphere as r increases is
\begin{eqnarray}\label{eq:K}
\m{dA}\;&=&\;\m{8\pi L^2\,a(t)\,[\, sinh^2(r/L)
da\;-\;{{dt}\over{L}}\, sinh(r/L)cosh(r/L)\,]} \nonumber \\
\;&=&\;\m{8\pi L\, sinh^2(r/L) \,a(t) \,dt\, [\,L
\dot{a}\;-\;coth(r/L)].}
\end{eqnarray}
For a trapped surface to exist in this spacetime, one must be able
to choose r and t in such a manner that dA is negative, that is, has
the same sign as dt. Now suppose that there is at least one value of
t, say t$^*$, such that L$\dot{\m{a}}$(t$^*$) $>$ 1 at that time; if
we assume the validity of the Einstein equations, then we see from
equation (\ref{eq:DENSITY}) that this is precisely equivalent to
assuming the existence of at least one spatial section on which
\emph{the total energy density is strictly positive}. Then the
spacelike hypersurface t = t$^*$ contains a trapped surface, because
if r is chosen sufficiently large then coth(r/L) [which of course is
always greater than unity, but approaches it as r tends to infinity]
becomes smaller than L$\dot{\m{a}}$(t$^*$).

For example, in the case where a bubble universe contains nothing
but pure positive vacuum energy, we obtain the version of de Sitter
spacetime with hyperbolic spatial sections, with the ``Bubble de
Sitter" metric:
\begin{eqnarray}\label{eq:R}
g(\mathrm{BdS})\; =\;
\m{-\,dt}^2\;+\;\m{sinh^2\big(t/L\big)}\,\Big[\mathrm{dr^2\;
+\;\m{L^2}\, sinh^2(r/L)}\{\mathrm{d}\theta^2 \;+\;
\mathrm{sin}^2(\theta)\,\mathrm{d}\phi^2\}\Big];
\end{eqnarray}
in this case we have
\begin{eqnarray}\label{eq:KK}
\m{dA}\;=\;\m{8\pi L\, sinh^2(r/L) \,sinh(t/L) \,dt\,
\Big[\,cosh(t/L)\;-\;coth(r/L)\Big].}
\end{eqnarray}
Notice that, in this case, it becomes steadily easier to keep the
expression in square brackets positive as time progresses [in the
sense that one need not take particularly large values of r in order
to ensure this]. The existence of trapped surfaces is in this sense
a local question at late stages of Inflation. Clearly, the spatial
volume does vanish at t = 0.

The Penrose theorem now explains why FRW spacetimes with negatively
curved spatial sections tend to be geodesically incomplete in the
past [since we are applying the theorem to past-directed null
geodesics]. All we needed were the very mild conditions that the NRC
should be satisfied ---$\,$ recall that this is equivalent to
assuming the NEC if the Einstein equations hold ---$\,$ and that
there should be at least one spatial section containing a trapped
set. Note that both of these conditions are satisfied by the version
of de Sitter spacetime with \emph{spherical} spatial sections. Thus,
contrary perhaps to intuition, what saves spatially spherical de
Sitter spacetime from being geodesically incomplete is \emph{not}
``gravitational repulsion" [that is, violation of the Strong Energy
Condition, which is not assumed in the Penrose theorem] but rather
the fact that the spatial sections do not have a non-compact
universal cover.

We conclude that the only way to avoid having a zero-volume
spacelike surface in a FRW bubble spacetime is to violate the NEC,
at least effectively.

This discussion used FRW geometry, so this result is not surprising;
but the advantage of using the Penrose theorem is that the argument
can be adapted to show that similar conclusions follow if an
\emph{inflating} bubble is perturbed, even to a large extent. Bubble
interiors do have spatial sections with non-compact universal
covers, and this \emph{topological} statement is robust against
perturbations. If we assume that Inflation occurs at late times,
then the existence of spatial surfaces with positive total energy
density at those times is only to be expected, since the positive
energy density of the inflaton will dominate; this will lead to the
existence of trapped surfaces. From another perspective: the
existence of trapped surfaces seems to be inevitable, since it is a
local question at late times, even in the case where the bubble has
been perturbed extensively at \emph{early} times. Thus, we expect to
be able to apply the Penrose theorem even to bubbles which are
\emph{not} close to a FRW form. It follows quite generally [even for
strongly perturbed spacetimes with highly irregular spatial
geometries] that \emph{an inflating bubble universe can only have an
``initial" spacelike surface with vanishing extrinsic curvature and
non-zero volume if the NEC is violated inside the bubble.}

A completely rigorous theory supporting this physical argument has
been given by Andersson and Galloway \cite{kn:andergall}, who prove
a theorem [Theorem 4.1] to the following effect. Suppose that we
take a globally hyperbolic asymptotically de Sitter spacetime
satisfying the NRC, and assume that the Cauchy surfaces [or their
universal covers] are \emph{not compact}. Suppose now that we try to
avoid having any spacelike surface with zero volume, by having a
bounce. [``Asymptotically de Sitter" is then assumed to hold both to
the past and to the future.] Then Andersson and Galloway show that
some future-directed null geodesic must fail to reach future
infinity. [The spacetime must also satisfy a certain genericity
condition, which essentially states that all spatial dimensions take
part in the accelerated expansion; see \cite{kn:according} for
further discussion, and see \cite{kn:pager} for another application
of results like this.] Since these spacetimes are supposed to evolve
to a de Sitter-like [inflationary] state [in which all
future-directed null geodesics do reach future infinity] we can
conclude that the NEC must indeed be violated by all spacetimes of
the kind in which we are interested here.

We now have an answer to our question as to how the matter content
of a bubble universe must be modified in order to avoid spacelike
sections of zero volume. The answer is simply that the NRC must be
violated inside the bubble, by some effect which is normally ignored
in discussions of bubble universes. This will involve either
modifying the Einstein equations so that the NRC can be violated
without violating the NEC, or directly violating the NEC itself. In
the next section, we explore the second option.

\addtocounter{section}{1}
\section* {\large{\textsf{4. Casimir Bubbles}}}

Our proposal is that the correct initial condition for a bubble
interior as it emerges from the quantum domain is that of vanishing
extrinsic curvature: this applies to the spacelike surface CD in
Figure 1. Formally, \emph{but not physically}, the geometry here is
like that of a ``bounce" cosmology \cite{kn:novello}; the great
difference is that, in our case, the initial conditions for the
semi-classical spacetime are \emph{not} prepared by an earlier
period of contraction. We stress that this is just a [natural]
\emph{proposal}: we have to verify that it makes sense physically,
within the context of string theory.

The idea that the \emph{Casimir effect} might play a crucial role in
cosmology has often been suggested: see for example
\cite{kn:coule}\cite{kn:godzong}\cite{kn:szydgod}\cite{kn:sahara}\cite{kn:levin}
and references therein. It has recently been raised in connection
with the ``standard model landscape" \cite{kn:nimah1}. As is well
known, the Casimir effect naturally leads to negative energy
densities and pressures, violating the NEC. This is of great
interest in string theory, because \emph{all} currently known
modulus stabilization schemes violate the NEC in one way or another.
[Furthermore, it seems likely \cite{kn:wesley} that NEC violation of
some kind is a fairly generic feature of theories involving higher
dimensions.] Subsequently \cite{kn:nimah2} it was found that by no
means all forms of NEC violation are acceptable in string theory;
Casimir effects are of great interest precisely because they belong
to the ``acceptable" class [outside the ``clock and rod" sector]. If
we wish to embed our discussion in string theory, then ``Casimir
cosmology" is a particularly natural ---$\,$ though surely not the
only ---$\,$ way to proceed.

The Casimir effect essentially arises from certain kinds of
\emph{boundary conditions} which one might find it physically
appropriate to impose. In the case of a bubble universe, we have to
ask: what kinds of boundary conditions are appropriate for fields
inside the bubble, and how can they be enforced?

This brings us directly to attempts [see particularly
\cite{kn:bouff} and references] to extend black hole complementarity
to cosmology. Recall that black hole complementarity resolves the
puzzles concerning Hawking radiation by declaring that one can
describe black hole radiation by taking either \emph{but not both}
of two points of view [following the star as it collapses or using
the observations of an observer who stays far away from it]. Either
perspective is postulated to give a complete description; paradoxes
only arise if one tries to take a ``global" point of view.

In the cosmological context, attention is focussed on \emph{causal
diamonds}, the entire region of a spacetime which is causally
connected to the worldline of a single observer. The remainder of
the global spacetime is then regarded as a set of redundant
descriptions of the same data, and, once again, paradoxes arise if
one attempts a global perspective. Now, in the case at hand, we wish
to apply this philosophy to the bubble universe portrayed in Figure
1. Take the observer whose worldline corresponds to the vertical
left-hand boundary of the diagram. The relevant part of the
corresponding causal diamond is represented by the dotted line. This
line intersects any spatial section [such as CD] at a \emph{finite}
distance from the observer. From the point of view of
complementarity, then, the spatial sections inside the bubble are
effectively finite; regarding them as infinite means taking the
``global" point of view of the bubble universe, and this is
precisely what complementarity forbids.

The problem of deciding how to implement this insight mathematically
is a difficult one. In order to proceed, we shall suggest a simple
ansatz, which is not intended to be fully realistic but which will
allow us to proceed in a quantitative way. Our suggestion is
prompted by the ideas discussed in References
\cite{kn:starr}\cite{kn:silvery}, in which the authors discuss
cosmological spacetimes with negatively curved spatial sections. As
is well known, it is possible to perform \emph{periodic
identifications} of domains in ordinary hyperbolic space H$^3$, so
that the quotient is compact. In classical general relativity this
makes no difference if the domain involved is very large, but in
\cite{kn:starr}\cite{kn:silvery} it is argued that string theory is
sensitive to such identifications, and that the periodic structure
has profound physical implications\footnote{Notice that the
\emph{local} isometry group of a compactified negatively curved
space is the same as that of ordinary hyperbolic space H$^3$, namely
O(1,3), since the local metric is completely unaffected by the
compactification. Therefore, the usual argument, whereby the O(4)
symmetry of the Euclidean instanton becomes the O(1,3) symmetry of
the spatial sections of the bubble, is unaffected.}.

Motivated by this, we propose to implement observer complementarity
in the following simple manner: when studying quantum-mechanical
aspects of the interior of a bubble universe, we should enforce
\emph{periodic boundary conditions} on all fields. Concretely, what
this idea means is that we should reject all fluctuations of fields
beyond a finite limit. Doing so will lead to a Casimir effect, which
will however be significant only in the very earliest era of the
bubble universe. We can now try to construct an internally
consistent model of an inflating bubble with spatial sections which,
with the help of Casimir energy, are able to avoid shrinking to size
zero at any time. Since we are interested in the very earliest
history of the bubble, where the inflaton is assumed to be rolling
extremely slowly, we can approximate the energy density of the
inflaton by that of a positive cosmological constant with
characteristic length scale L; the negative Casimir energy density
is superimposed on this.

Casimir energies can depend sensitively on the kinds of matter
fields involved and whether the effects of higher dimensions are to
be taken into account, and so on; but let us continue to proceed in
the simplest possible manner, and assume as usual \cite{kn:levin}
that, for a four-dimensional FRW spacetime with effectively compact
spatial sections, the Casimir density depends on the inverse fourth
power of the scale factor. The total energy density is then a
combination of the background vacuum density $+\,3/8\pi$L$^2$ with
the Casimir energy; so the Friedmann equation takes the form
\begin{eqnarray}\label{eq:T}
\m{L^2\,\dot{a}^2\;=\;{{8\pi}\over{3}}\,L^2\,a^2\Big[{{3}\over{8\pi
L^2}}\;-\;{{6}\over{8\pi L^2\,a^4}}\Big]\;+\;1.}
\end{eqnarray}
Here the coefficient of the Casimir term has been fixed by requiring
the surface of zero extrinsic curvature to correspond to a scale
factor equal to unity. The solution for the ``Bubble de Sitter plus
Casimir" metric is remarkably simple:
\begin{eqnarray}\label{eq:U}
g(\m{BdS+C}) = \m{-\;dt^2 + \Big[
\,1\;+\;3\,sinh^2(t/L)\Big]\Big[dr^2 + \m{L^2}\,
sinh^2(r/L)}\{\mathrm{d}\theta^2 +
\mathrm{sin}^2(\theta)\,\mathrm{d}\phi^2\}\Big].
\end{eqnarray}
Notice that this is asymptotic, as t tends to infinity, to Bubble
de Sitter spacetime [equation (\ref{eq:R}); the factor of 3 can be
absorbed in the limit], but it has a spatial surface of zero
extrinsic curvature at t = 0. If we simply postulate that the
semi-classical bubble history begins at that time, then we have a
picture of the bubble interior in which the Casimir effect is
significant for a very brief period, which is succeeded [as the
Casimir energy rapidly dilutes but the inflaton energy does not]
by an ordinary accelerated expansion. The Casimir effect allows
the low-entropy conditions in the exterior to establish, via the
surface t = 0, similar conditions in the interior; having done
this duty, it rapidly disappears, and the usual description of a
bubble interior becomes valid.

The Casimir effect is completely harmless at the perturbative level,
but it is far from clear that this remains true non-perturbatively,
particularly when it plays such an important role in fixing the
spacetime geometry. In fact, it is known that such effects can lead
to serious consequences \cite{kn:singularstable}, as follows.
Seiberg and Witten \cite{kn:seiberg} observed that branes, being
extended objects, can be extremely sensitive to the geometry of the
spaces in which they propagate. If the geometry takes certain forms,
it can actually lead to a situation which Maldacena and Maoz
\cite{kn:maoz} [see also \cite{kn:porrati}] describe as a
\emph{pair-production instability} for branes.

To be specific: suppose that a given spacetime has a Euclidean
version which is conformally compactifiable; that is, it is
conformal to the interior of a compact manifold-with-boundary. Such
manifolds are said to be \emph{asymptotically hyperbolic}: that is,
the geometry comes to resemble that of hyperbolic space\footnote{In
our case this space will be four-dimensional; it should not be
confused with the [also hyperbolic] three-dimensional transverse
slices of the Lorentzian version.} at sufficiently large distances.
For Euclidean BPS branes in four dimensions, the brane action
consists of two terms: a positive one proportional to the
[three-dimensional] area of the brane, and a negative one
proportional to the volume enclosed by it. So we have, in four
dimensions,
\begin{equation}\label{eq:V}
\mathrm{S} \;=\;
\Theta(\mathrm{A}\;-\;{{\mathrm{3}}\over{\mathrm{L}}}\,\mathrm{V}),
\end{equation}
where $\Theta$ is the tension, A is the area, V the volume enclosed,
and L is the background asymptotic curvature radius. If at any point
the volume term is larger than the area term, it will be possible to
reduce the action of the system by creating brane-antibrane pairs
and moving them to the appropriate positions, as described by
Maldacena and Maoz \cite{kn:maoz}. Thus a severe non-perturbative
instability will arise. In this way we obtain a powerful criterion
for the acceptability of specific geometries from a stringy point of
view: powerful because it applies even when the NEC is only violated
\emph{effectively}.

To see how this works in the present case, let us proceed as
follows. We begin by constructing the asymptotically hyperbolic
version of Bubble de Sitter spacetime, with metric given in equation
(\ref{eq:R}). We simply complexify both t and L, but not r.
Re-labelling the latter as L$\chi$, we obtain
\begin{equation}\label{eq:RR}
g(\m{AHBdS)\; =\; dt^2\;+\;L^2\,sinh^2\big(t/L\big)\,\Big[d\chi^2\;
+\;sin^2(\chi)\{d\theta^2 \;+\; sin^2(\theta)\,d\phi^2}\}\Big];
\end{equation}
this ``Asymptotically Hyperbolic Bubble de Sitter" metric is in fact
the metric of four-dimensional hyperbolic space, foliated by
three-spheres. Note that the sign of the curvature has been reversed
by the complexification of L. [In order to obtain \emph{anti-de
Sitter} spacetime from H$^4$, one chooses a quite different
foliation, with negatively curved slices, and of course one does not
complexify L; see \cite{kn:OVV} for the details.] Notice that this
foliation makes it obvious that the conformal boundary is positively
curved; this is important for establishing non-perturbative
stability at \emph{large} values of t, as was shown by Seiberg and
Witten \cite{kn:seiberg}. In fact, the brane action in this case can
be evaluated explicitly: from equation (\ref{eq:V}) we have
\begin{equation}\label{eq:RRR}
\m{S[BdS](t)\;=\;2\pi^2\,\Theta\,L^3\,\Big[sinh^3(t/L)\;-\;{{1}\over{4}}\,cosh(3t/L)\;+\;{{9}\over{4}}\,cosh(t/L)\;-\;2\,
\Big].}
\end{equation}
This function is actually non-negative at \emph{all} positive values
of t, large or small, so Bubble de Sitter spacetime is
\emph{completely stable} against this particular non-perturbative
effect. Actually, the function increases monotonically with t; this
is characteristic of spatially flat or negatively curved
asymptotically de Sitter spacetimes which satisfy the NEC. When the
NEC is violated, there are grounds for serious concern that the
action will not behave so benignly.

Applying this same complexification to the metric in equation
(\ref{eq:U}), we have the asymptotically hyperbolic version of
$g(\m{BdS+C})$:
\begin{eqnarray}\label{eq:UU}
g(\m{AHBdS+C})= \m{dt^2\;+\;L^2\,\Big[
\,1\;+\;3\,sinh^2(t/L)\Big]\Big[\,d\chi^2
  +  \,sin^2(\chi)}\{\mathrm{d}\theta^2 +
\mathrm{sin}^2(\theta)\,\mathrm{d}\phi^2\}\Big].
\end{eqnarray}
Note that t/L is not complexified, so we can still interpret it as a
dimensionless measure of time in this case. If we truncate this
space at t = T, then the brane action for t $\geq$ T is
\begin{eqnarray}\label{eq:Y}
\m{S[BdS+C]_{T}(t)}\; &=& \;\m{2\pi^2\,\Theta\,L^3\,\Big[\,\Big(
\,1\;+\;3\,sinh^2(t/L)\Big)^{3/2}\;} \nonumber \\
                        &-& \;\m{{{3}\over{L}}\,\int_{T}^t\,\Big(
\,1\;+\;3\,sinh^2(\tau/L)\Big)^{3/2}d\tau\Big]},
\end{eqnarray}
where $\Theta$ is the tension, as in equation (\ref{eq:V}).

If T = 0, this function begins at t = 0 with a positive value equal
to $2\pi^2\,\Theta\,$L$^3$ and then immediately \emph{declines} as t
increases. This decrease is characteristic of NEC-violating
spacetimes, as was shown in the case of \emph{flat} compact spatial
sections in \cite{kn:singularstable}; it is the reason for the fact
that NEC-violating spacetimes are in danger of being
non-perturbatively unstable. The positive curvature of the t =
constant sections in this case means that
---$\,$ as in the case of Bubble de Sitter space ---$\,$ there
is no such instability at \emph{large} values of t, but there might
be a problem at \emph{small} values of t if the NEC violation causes
the action to fall too low. [Maldacena and Maoz \cite{kn:maoz}
discuss examples where this happens.] Since the Seiberg-Witten
argument shows that the action \emph{is} positive at large t, its
decline must be halted at some point. The question is whether it is
halted in time to prevent the action from becoming negative.

\begin{figure}[!h] \centering
\includegraphics[width=0.7\textwidth]{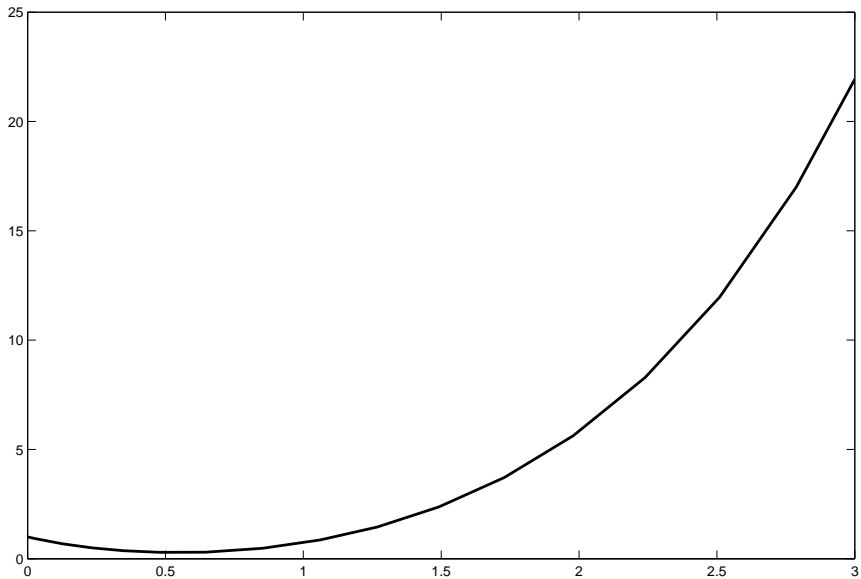}
\caption{The Action $\m{S[BdS+C]_{0}(t)}$.}
\end{figure}

The graph of the action function has a unique minimum [for all T] at
t = (ln($\sqrt{3}$))L. In the case where we cut off the space at its
neck [so that the spatial sections inside the bubble never contract,
which is what we are supposing here], we have T = 0, and a simple
numerical investigation shows that the action decreases from a
positive value at t = 0 down to S[BdS+C]$_0$((ln($\sqrt{3}$))L),
which is still \emph{positive} [for all L]. The brane action
subsequently increases indefinitely as the area term decisively
overcomes the volume term as the Casimir energy is diluted [so that
the action function comes to resemble that of Bubble de Sitter].
Thus the action is positive everywhere. Given this, it is easy to
see that the same statement holds true for any T $\geq$ 0:
\emph{there is no Seiberg-Witten instability in this system, as long
as the spatial sections never contract.} The graph\footnote{The
horizontal axis is t/L, the vertical axis is
$\m{S[BdS+C]_{0}(t)/2\pi^2\,\Theta\,L^3}$.} of the action for T = 0
is given in Figure 3; notice that the system escapes from being
unstable despite the initial decrease of the action.

Allowing the spatial sections to contract means taking T to be
negative. In this case, the initial value of S[BdS+C]$_{\m{T}}$(t)
becomes a larger positive number; but on the other hand the function
decreases for a longer time, so it is not obvious that it remains
positive everywhere. A numerical investigation shows that, as T is
modified downwards, S[BdS+C]$_{\m{T}}$((ln($\sqrt{3}$))L), the
minimum value of the action, stays non-negative only down to a value
of T that is very close to zero, T $\approx$ $-$0.0928L. The scale
factor at that value of t is given approximately by a($-$0.0928L)
$\approx$ 1.00644. Clearly there is essentially no contraction in
this case [the minimum value of the scale factor being unity].

We interpret this last result as strong evidence in favour of our
postulate that the bubble history begins on or near to the surface
of vanishing extrinsic curvature: if it tries to begin earlier, the
system becomes violently unstable. The spacetime geometry is
\emph{not} like that of a ``bounce" spacetime: there is little or no
contraction as seen by the distinguished bubble observers.

Of course, the example we have considered here is a very special
one: it is motivated by a desire to present a fully explicit metric.
In fact, Casimir effects are not the only way to achieve NEC
violation [or ``effective" NEC violation
---$\,$ see \cite{kn:singularstable}\cite{kn:stachszyd}]. However, numerical
experiments lead us to believe that if the NEC is violated,
effectively or otherwise, in ways that are compatible with the ideas
of Arkani-Hamed et al \cite{kn:nimah2}, then one will be led to a
picture similar to the one presented in detail here: that is, the
requirement of non-perturbative string stability will prohibit any
more than a negligible amount of contraction inside a bubble
universe.

In summary, it is very difficult for a bubble universe to resemble
our world, because to do so it needs to begin with very special and
delicate properties; but it may be possible if NEC violation is
indeed compatible with, \emph{yet constrained by}, stringy
considerations.

\addtocounter{section}{1}
\section* {\large{\textsf{5. Conclusion: Building a Landscape}}}

In the stringy picture of ``creation from nothing" [or the
``emergence of time"] \cite{kn:ooguri}, the original ``mother of all
universes" is born along a spatial section that is as smooth as it
can be, up to quantum fluctuations \cite{kn:arrow}\cite{kn:BBB}.
This allows Inflation to start in the mother universe. The latter
may however subsequently nucleate bubble universes of the kind we
have been considering in this work. The Arrow in these bubbles, if
any, must be \emph{inherited} from the mother universe; the Arrow
can then be handed down to subsequent generations. In this way we
obtain an explanation of the observed Arrow that does not involve
wildly improbable or rare fluctuations into lower-entropy states. In
this work, we have suggested a way of ensuring that this process of
``inheritance" does occur.

However, the argument in favour of ``Arrow inheritance" in the
NEC-violating case does depend on the ``causal diamond" or
``observer complementarity" philosophy. We needed this principle
to justify the compactification of the bubble's spatial sections
---$\,$ or ``periodic boundary conditions"
---$\,$ used in the previous section. While this idea is well
motivated by black hole complementarity [and by ideas from string
theory \cite{kn:starr}\cite{kn:silvery}], the extrapolation to
cosmological horizons is not entirely secure. We should therefore
ask: what would be the consequences for the landscape if this
extrapolation had to be abandoned?

In that case, we would be led to conclude that \emph{we are in the
original universe}, the one presented to us directly by creation
from ``nothing". For this original universe does have an Arrow of
time, such as we in fact observe; whereas no bubble universe would
have this remarkable property. This would drastically change the
role of bubble universes: far from seeding new life, they would
merely destroy any ordered structure with which they collided in the
original universe. This phenomenon might have to be taken into
account in discussions of the nature of observers at very late
times.

\emph{If} bubble universes are unable to inherit an Arrow, then we
must find another way of building a Landscape ---$\,$ that is, of
actually constructing universes which realise the full set of string
vacuum solutions. A way of doing so which automatically gives rise
to spacetimes with low initial entropy is suggested by the work of
Gibbons and Hartle \cite{kn:gibhart}, who raised the interesting
question as to whether a universe created from ``nothing" must be
topologically \emph{connected.} This is not at all obvious, because
a compact manifold-with-boundary can, and generically will, have a
boundary which breaks up into disconnected pieces; this idea is
familiar from \emph{cobordism} theory \cite{kn:lawson}. Gibbons and
Hartle gave an elegant proof that the boundary must indeed be
connected in the Hartle-Hawking case if all eigenvalues of the Ricci
curvature of the Euclidean space are positive and bounded away from
zero. This condition is certainly not satisfied by the spaces used
in the work of Ooguri et al., however, and so the question remains
open. If indeed the relevant Euclidean space has multiple boundary
components of zero extrinsic curvature, then potentially large
numbers of spacetimes can be born from a single Euclidean ancestor;
those born from a boundary component with a suitable [toral]
topology will have an Arrow, as explained in \cite{kn:arrow}. The
question then, of course, will be whether these universes have
suitably spaced values of the cosmological constant. Perhaps the
methods of Dijkgraaf et al. \cite{kn:goodbaby} can be adapted to
study this.

The Lorentzian spacetimes so created would be completely mutually
inaccessible. However, it might be possible to find indirect
evidence of the existence of the other universes in our own past,
since all universes originate from a common Euclidean space.

\addtocounter{section}{1}
\section*{\large{\textsf{Acknowledgements}}}
The author is grateful to Dr. Soon Wanmei for the diagrams, and for
her unfailing care and concern. He would also like to thank Matthew
Kleban for patient and extremely helpful correspondence, and the
referee whose questions and suggestions have helped to clarify the
argument at many points.


\begin{thebibliography}{18}
\bibitem{kn:landscape}
Leonard Susskind, The Anthropic Landscape of String Theory,
arXiv:hep-th/0302219
\bibitem{kn:deluccia}
Sidney R. Coleman, Frank De Luccia, Gravitational Effects On And Of
Vacuum Decay, Phys.Rev. D21 (1980) 3305
\bibitem{kn:frei}
Ben Freivogel, Matthew Kleban, Maria Rodriguez Martinez, Leonard
Susskind, Observational Consequences of a Landscape, JHEP 0603
(2006) 039, \x hep-th/0505232
\bibitem{kn:kleban}
Puneet Batra, Matthew Kleban, Transitions Between de Sitter Minima,
Phys. Rev. D 76 (2007) 103510, \x hep-th/0612083
\bibitem{kn:lindepotential}
Andrei Linde, Misao Sasaki, Takahiro Tanaka, CMB in Open Inflation,
Phys.Rev. D59 (1999) 123522, arXiv:astro-ph/9901135
\bibitem{kn:trodden}
Tanmay Vachaspati, Mark Trodden, Causality and Cosmic Inflation,
Phys.Rev. D61 (2000) 023502, \x gr-qc/9811037
\bibitem{kn:albrecht}
Andreas Albrecht, Cosmic Inflation and the Arrow of Time, in
\emph{Science and Ultimate Reality: Quantum Theory, Cosmology and
Complexity}, eds J. D. Barrow, P.C.W. Davies, C.L. Harper, Cambridge
University Press (2004), \x astro-ph/0210527
\bibitem{kn:couleinitial}
D. H. Coule, Difficulties with inflationary initial conditions,
arXiv:0706.0205

\bibitem{kn:penrose}
R. Penrose, Singularities and Time-Asymmetry, in \emph{General
Relativity: An Einstein Centenary Survey}, eds S W Hawking, W
Israel, Cambridge University Press, 1979

\bibitem{kn:price}
Huw Price, Cosmology, Time's Arrow, and That Old Double Standard, in
\emph{Time's Arrows Today}, ed S. Savitt, Cambridge University Press
1994, \x gr-qc/9310022; The Thermodynamic Arrow: Puzzles and
Pseudo-puzzles, \x physics/0402040
\bibitem{kn:dyson}
Lisa Dyson, Matthew Kleban, Leonard Susskind, Disturbing
Implications of a Cosmological Constant, JHEP 0210 (2002) 011, \x
hep-th/0208013
\bibitem{kn:wald}
Robert M. Wald, The Arrow of Time and the Initial Conditions of the
Universe, \x gr-qc/0507094
\bibitem{kn:carroll}
Sean M. Carroll, Jennifer Chen, Spontaneous Inflation and the Origin
of the Arrow of Time, \x hep-th/0410270
\bibitem{kn:vilenkin}
A. Vilenkin, Creation of Universes from Nothing, Phys.Lett.B117
(1982) 25
\bibitem{kn:ooguri}
Hirosi Ooguri, Cumrun Vafa, Erik Verlinde, Hartle-Hawking
Wave-Function for Flux Compactifications: The Entropic Principle,
Lett.Math.Phys. 74 (2005) 311, \x hep-th/0502211

\bibitem{kn:black}
Paul Frampton, Stephen D.H. Hsu, Thomas W. Kephart, David Reeb, What
is the entropy of the universe? arXiv:0801.1847 (hep-th)
\bibitem{kn:BBB}
Brett McInnes, The Arrow of Time in the Landscape, arXiv:0711.1656
[hep-th]
\bibitem{kn:aguirregrat2}
Anthony Aguirre, Steven Gratton, Inflation without a beginning: a
null boundary proposal, Phys.Rev. D67 (2003) 083515,
arXiv:gr-qc/0301042
\bibitem{kn:arrow}
Brett McInnes, Arrow of Time in String Theory, Nucl. Phys. B782
(2007) 1, \x hep-th/0611088
\bibitem{kn:laura}
Claus Kiefer, Braz.J.Phys. 35 (2005) 296, arXiv:gr-qc/0502016; R.
Holman, L.Mersini-Houghton, \x hep-th/0511102, Phys.Rev. D74 (2006)
123510, \x hep-th/0512070; A.Gorsky, Phys.Lett.B646 (2007) 183, \x
hep-th/0606072; Matias Aiello, Mario Castagnino, Olimpia Lombardi,
\x gr-qc/0608099; Laura Mersini-Houghton, \x gr-qc/0609006
\bibitem{kn:banks}
T. Banks, Entropy and initial conditions in cosmology, \x
hep-th/0701146
\bibitem{kn:freese}
Katherine Freese, Matthew G. Brown, William H. Kinney, The Phantom
Bounce: A New Proposal for an Oscillating Cosmology, arXiv:0802.2583
(astro-ph)
\bibitem{kn:uzan}
Cyril Pitrou, Thiago S. Pereira, Jean-Philippe Uzan, Predictions
from an anisotropic inflationary era, arXiv:0801.3596 [astro-ph]




\bibitem{kn:borde}
Arvind Borde, Alex Vilenkin, Phys.Rev. D56 (1997) 717, \x
gr-qc/9702019; Arvind Borde, Alan H. Guth, Alexander Vilenkin,
Phys.Rev.Lett. 90 (2003) 151301, \x gr-qc/0110012
\bibitem{kn:guthnew}
Alan H. Guth, Eternal inflation and its implications, J.Phys.A40
(2007) 6811, \x hep-th/0702178
\bibitem{kn:aguirrevaas}
Anthony Aguirre, Eternal Inflation, past and future, arXiv:0712.0571
\bibitem{kn:aguirregrat}
Anthony Aguirre, Steven Gratton, Steady-State Eternal Inflation,
Phys.Rev. D65 (2002) 083507, arXiv:astro-ph/0111191
\bibitem{kn:aguirrejoh}
Anthony Aguirre, Matthew C Johnson, Towards observable signatures of
other bubble universes II: Exact solutions for thin-wall bubble
collisions, arXiv:0712.3038


\bibitem{kn:hsu}
Roman V. Buniy, Stephen D.H. Hsu, A. Zee, Does string theory predict
an open universe? Phys.Lett.B660 (2008) 382, \x hep-th/0610231
\bibitem{kn:bojowald}
Martin Bojowald, Reza Tavakol, Loop Quantum Cosmology: Effective
theories and oscillating universes arXiv:0802.4274
\bibitem{kn:unstable}
Brett McInnes, The Phantom Divide in String Gas Cosmology,
Nucl.Phys. B718 (2005) 55, \x hep-th/0502209
\bibitem{kn:tallandthin}
Brett McInnes, The Most Probable Size of the Universe, Nucl.Phys.
B730 (2005) 50, \x hep-th/0509035
\bibitem{kn:singularstable}
Brett McInnes, Pre-Inflationary Spacetime in String Cosmology, Nucl.
Phys. B748 (2006) 309, \x hep-th/0511227
\bibitem{kn:ovrut}
Evgeny I. Buchbinder, Justin Khoury, Burt A. Ovrut, New Ekpyrotic
Cosmology, Phys. Rev. D76 (2007) 123503, \x hep-th/0702154; On the
Initial Conditions in New Ekpyrotic Cosmology, JHEP 11 (2007) 076,
arXiv:0706.3903
\bibitem{kn:cremi}
Paolo Creminelli, Leonardo Senatore, A smooth bouncing cosmology
with scale invariant spectrum, JCAP 0711:010,2007, \x hep-th/0702165
\bibitem{kn:carrtrod}
Sean M. Carroll, Mark Hoffman, Mark Trodden, Can the dark energy
equation-of-state parameter w be less than $-$1?, Phys.Rev. D68
(2003) 023509, arXiv:astro-ph/0301273

\bibitem{kn:nimah1}
Nima Arkani-Hamed, Sergei Dubovsky, Alberto Nicolis, Giovanni
Villadoro, Quantum Horizons of the Standard Model Landscape, JHEP 06
(2007) 078, arXiv:hep-th/0703067
\bibitem{kn:nimah2}
Nima Arkani-Hamed, Sergei Dubovsky, Alberto Nicolis, Enrico
Trincherini, Giovanni Villadoro, A Measure of de Sitter Entropy and
Eternal Inflation, JHEP 0705 (2007) 055, arXiv:0704.1814 [hep-th]
\bibitem{kn:coule}
D.H. Coule, Quantum Cosmological Models, Class.Quant.Grav. 22 (2005)
R125, \x gr-qc/0412026

\bibitem{kn:seiberg}
Nathan Seiberg, Edward Witten, The D1/D5 System And Singular CFT,
JHEP 9904 (1999) 017, \x hep-th/9903224
\bibitem{kn:maoz}
Juan Maldacena, Liat Maoz, Wormholes in AdS, JHEP 0402 (2004) 053,
\x hep-th/0401024

\bibitem{kn:porrati}
M. Kleban, M. Porrati, R. Rabadan, Stability in Asymptotically AdS
Spaces, JHEP 0508 (2005) 016, \x hep-th/0409242


\bibitem{kn:fargu}
Edward Farhi, Alan H. Guth, An Obstacle To Creating A Universe In
The Laboratory, Phys.Lett.B183 (1987) 149-155
\bibitem{kn:aguirrejohn}
Anthony Aguirre, Matthew C. Johnson, Two Tunnels to Inflation,
Phys.Rev. D73 (2006) 123529, arXiv:gr-qc/0512034

\bibitem{kn:tod}
Filipe C. Mena, Paul Tod, Lanczos potentials and a definition of
gravitational entropy for perturbed FLRW space-times,
Class.Quant.Grav.24 (2007) 1733, \x gr-qc/0702057
\bibitem{kn:ram}
Ram Brustein, Cosmological Entropy Bounds, \x hep-th/0702108
\bibitem{kn:gibhawk}
G.W. Gibbons, S.W. Hawking, Cosmological Event Horizons,
Thermodynamics, and Particle Creation, Phys.Rev.D15 (1977), 2738


\bibitem{kn:gibsol}
G. W. Gibbons, S. N. Solodukhin, The Geometry of Large Causal
Diamonds and the No Hair Property of Asymptotically de-Sitter
Spacetimes, Phys.Lett.B652 (2007) 103, arXiv:0706.0603


\bibitem{kn:turok}
Joel K. Erickson, Daniel H. Wesley, Paul J. Steinhardt, Neil Turok,
Kasner and Mixmaster behavior in universes with equation of state w
$\geq$ 1, Phys.Rev. D69 (2004) 063514, \x hep-th/0312009
\bibitem{kn:uggla}
J. Mark Heinzle, Claes Uggla, Niklas Rohr, The cosmological billiard
attractor, \x gr-qc/0702141
\bibitem{kn:imponente}
Giovanni Montani, Marco Valerio Battisti, Riccardo Benini, Giovanni
Imponente, Classical and Quantum Features of the Mixmaster
Singularity, arXiv:0712.3008 [gr-qc]
\bibitem{kn:linde}
Andrei Linde, Sinks in the Landscape, Boltzmann Brains, and the
Cosmological Constant Problem, JCAP 0701 (2007) 022, \x
hep-th/0611043
\bibitem{kn:page}
Don N. Page, Is Our Universe Decaying at an Astronomical Rate?, \x
hep-th/0612137
\bibitem{kn:carlip}
S. Carlip, Transient Observers and Variable Constants, or Repelling
the Invasion of the Boltzmann's Brains, JCAP 0706 (2007) 001, \x
hep-th/0703115
\bibitem{kn:gott} J. Richard Gott III, Boltzmann
Brains--I'd Rather See Than Be One, arXiv:0802.0233 (gr-qc)

\bibitem{kn:podolsky2}
D. Podolsky, General asymptotic solutions of the Einstein equations
and phase transitions in quantum gravity, arXiv:0704.0354 [hep-th]
\bibitem{kn:gagv}
Jaume Garriga, Alan H. Guth, Alexander Vilenkin, Eternal inflation,
bubble collisions, and the persistence of memory, Phys.Rev.D76
(2007) 123512, \x hep-th/0612242
\bibitem{kn:freihorshe}
Ben Freivogel, Gary T. Horowitz, Stephen Shenker, Colliding with a
Crunching Bubble, JHEP 0705 (2007) 090, arXiv:hep-th/0703146
\bibitem{kn:aguijohsho}
Anthony Aguirre, Matthew C Johnson, Assaf Shomer, Towards observable
signatures of other bubble universes, Phys.Rev.D76 (2007) 063509,
arXiv:0704.3473
\bibitem{kn:fish}
Willy Fischler, Chethan Krishnan, Sonia Paban, Marija Zanic, Vacuum
Bubble in an Inhomogeneous Cosmology: A Toy Model, arXiv:0711.3417
[hep-th]
\bibitem{kn:worldscollide}
Spencer Chang, Matthew Kleban, Thomas S. Levi, When Worlds Collide,
arXiv:0712.2261
\bibitem{kn:strominger}
Andrew Strominger, David Thompson, A Quantum Bousso Bound, Phys.Rev.
D70 (2004) 044007, arXiv:hep-th/0303067
\bibitem{kn:bouff}
Raphael Bousso, Ben Freivogel, A paradox in the global description
of the multiverse, JHEP 0706 (2007) 018 arXiv:hep-th/0610132
\bibitem{kn:verlinde}
Erik Verlinde, On the Holographic Principle in a Radiation Dominated
Universe, arXiv:hep-th/0008140
\bibitem{kn:huang1}
Qing-Guo Huang, Observational Consequences of Quantum Cosmology,
Nucl.Phys.B777 (2007) 253, arXiv:hep-th/0510219
\bibitem{kn:huang2}
Qing-Guo Huang, Holographic Principle and Quantum Cosmology,
arXiv:hep-th/0512004
\bibitem{kn:novello}
M. Novello, S.E.Perez Bergliaffa, Bouncing Cosmologies,
arXiv:0802.1634 (astro-ph)
\bibitem{kn:piao}
Yi-Fu Cai, Taotao Qiu, Yun-Song Piao, Mingzhe Li, Xinmin Zhang,
Bouncing Universe with Quintom Matter, JHEP 0710 (2007) 071,
arXiv:0704.1090
\bibitem{kn:peter}
Felipe T. Falciano, Marc Lilley, Patrick Peter, A classical bounce:
constraints and consequences, arXiv:0802.1196 (gr-qc)

\bibitem{kn:couleholo}
D.H. Coule, Holography constrains quantum bounce, arXiv:0802.1867
(gr-qc)


\bibitem{kn:tiplerreview}
Tipler, F. J., Clarke, C. J. S., Ellis, G. F. R., Singularities
 and horizons --- a review article. In  \emph{General relativity and gravitation, Vol. 2}, Plenum, 1980

\bibitem{kn:waldbook}
Robert M. Wald, \emph{General Relativity}, Chicago University Press,
1984


\bibitem{kn:andergall}
L. Andersson, G.J. Galloway, dS/CFT and spacetime topology,
Adv.Theor.Math.Phys. 6 (2003) 307, arXiv:hep-th/0202161
\bibitem{kn:according}
Brett McInnes, De Sitter and Schwarzschild-De Sitter According to
Schwarzschild and De Sitter, JHEP 09(2003)009, arXiv:hep-th/0308022
\bibitem{kn:pager}
Don N. Page, No-bang quantum state of the holocosm, arXiv:0707.2081


\bibitem{kn:godzong}
Wlodzimierz Godlowski, Marek Szydlowski, Zong-Hong Zhu, Constraining
bouncing cosmology caused by Casimir effect, Gravitation and
Cosmology 14 (2008) 17, arXiv:astro-ph/0702237
\bibitem{kn:szydgod}
Marek Szydlowski, Wlodzimierz Godlowski, Acceleration of the
Universe driven by the Casimir force, arXiv:0705.1772
\bibitem{kn:sahara}
A. A. Saharian, M. R. Setare, Casimir effect in de Sitter spacetime
with compactified dimension, Phys. Lett. B 659 (2008) 367,
arXiv:0707.3240
\bibitem{kn:levin}
Brian R. Greene, Janna Levin, Dark Energy and Stabilization of Extra
Dimensions, JHEP 0711:096,2007, arXiv:0707.1062 [hep-th]
\bibitem{kn:wesley}
Daniel H. Wesley, Oxidised cosmic acceleration, arXiv:0802.3214
(hep-th)
\bibitem{kn:starr}
John McGreevy, Eva Silverstein, David Starr, New Dimensions for
Wound Strings: The Modular Transformation of Geometry to Topology,
Phys.Rev. D75 (2007) 044025, arXiv:hep-th/0612121
\bibitem{kn:silvery}
Daniel Green, Albion Lawrence, John McGreevy, David R. Morrison, Eva
Silverstein, Dimensional Duality, Phys.Rev.D76 (2007) 066004,
arXiv:0705.0550 [hep-th]
\bibitem{kn:OVV}
Brett McInnes, The Geometry of The Entropic Principle and the Shape
of the Universe, JHEP 10 (2006) 029, \x hep-th/0604150
\bibitem{kn:stachszyd}
Tomasz Stachowiak, Marek Szydlowski, Exact solutions in bouncing
cosmology, Phys.Lett. B646 (2007) 209, arXiv:gr-qc/0610121

\bibitem{kn:gibhart}
G.W. Gibbons, J.B. Hartle, Real Tunneling Geometries and the
Large-Scale Topology of the Universe, Phys.Rev. D42 (1990) 2458
\bibitem{kn:lawson}
H. Blaine Lawson and Marie-Louise Michelsohn, \emph{Spin Geometry},
Princeton University Press, Princeton, 1990
\bibitem{kn:goodbaby}
Robbert Dijkgraaf, Rajesh Gopakumar, Hirosi Ooguri, Cumrun Vafa,
Baby Universes in String Theory, Phys.Rev. D73 (2006) 066002, \x
hep-th/0504221























\end{thebibliography}
\end{document}